\documentclass[CEJP,DVI]{cej} 
\usepackage{layout}
\usepackage{amsmath}
\usepackage{textcomp}
\usepackage{hyperref}

\title{Suppression of the repulsive force in nuclear interactions
near the chiral phase transition}

\articletype{Short Communication}

\author{Chihiro Sasaki\email{sasaki@fias.uni-frankfurt.de}}

\institute{
Frankfurt Institute for Advanced Studies,
D-60438 Frankfurt am Main,
Germany
          }

\abstract{
We introduce an effective chiral Lagrangian with a dilaton 
responsible for the trace anomaly in QCD.
As the ``dilaton limit'' is taken, which drives a system to near 
chiral restoration density, a linear sigma model emerges from
the highly non-linear structure. 
A striking prediction is that the vector-meson--nucleon interaction 
gets strongly suppressed when the dilaton limit is approached.
Its phenomenological implications for the thermodynamics of
dense baryonic matter are briefly discussed.
}


\begin{document}
\maketitle

\section{Role of the dilaton at high density}

In the limit of massless quarks 
the QCD Lagrangian possesses the chiral symmetry and scale invariance, both
of which are dynamically broken in the physical vacuum due to the strong 
interaction. The QCD trace anomaly signals the emergence of a scale at the
quantum level from the theory without any dimension-full parameters.
Therefore, spontaneous chiral symmetry breaking, which gives rise to a nucleon
mass, and the trace anomaly have an intimate connection to each other and 
dynamical scales in hadronic systems are considered to originate from them.
In nuclear physics, a scalar meson plays an essential role
as known from Walecka model that works fairly well for phenomena near nuclear
matter density~\cite{walecka}. On the other hand, at high density, the relevant 
Lagrangian possessing correct symmetry is the linear sigma model, and the scalar 
needed there is the fourth component of the chiral four-vector $(\vec{\pi},\sigma)$.
Thus in order to probe highly dense matter, we have to resolve how the
chiral scalar at low density transmutes to the fourth component
of the four-vector.

The trace anomaly is implemented in a chiral Lagrangian by introducing
a dilaton (or glueball) field representing the gluon condensate 
$\langle G_{\mu\nu}G^{\mu\nu} \rangle$~\cite{schechter}. 
Following~\cite{miransky}, we express the trace anomaly in terms of ``soft'' 
$\chi_s$ and ``hard'' $\chi_h$ dilatons.  
We associate the soft dilaton with that component locked to the quark 
condensate $\langle\bar{q}q\rangle$~\cite{LeeRho}. This is assumed to be the 
component which ``melts'' across the chiral phase transition whereas the hard 
one remains non-vanishing.
The soft dilaton plays an important role in the emergence of a half-skyrmion 
phase at high density where a skyrmion turns into two half-skyrmions~\cite{half}.

In introducing baryons, there are two alternative ways of 
assigning chirality to the nucleons, ``naive'' and mirror assignments. 
The ``naive'' assignment is anchored on 
the standard chiral symmetry structure where the entire constituent quark or 
nucleon mass (in the chiral limit) is generated by spontaneous symmetry breaking.
The alternative, mirror assignment~\cite{dk,mirror}
allows a chiral invariant mass term
which remains non-zero at chiral restoration.
Therefore a part of the nucleon mass, $m_0$, must arise 
from a mechanism that is not associated with spontaneous chiral symmetry breaking. 
The origin of such a mass $m_0$ can be traced back to the non-vanishing
gluon condensate in chiral symmetric phase and the broken scale 
symmetry is accounted for by the hard dilaton.
In this way the origin of $m_0$ is attributed to the hard component of the gluon 
condensate~\cite{SLPR}.

\section{Dilaton limit}

Our effective Lagrangian for the Nambu-Goldstone bosons, vector mesons and 
soft dilatons is derived~\cite{SLPR} starting with 
the hidden local symmetric (HLS)~\cite{hls} Lagrangian following the strategy 
of Beane and van Kolck~\cite{vanKolck}.
Conformal invariance can be embedded in chiral Lagrangians by
introducing a scalar field $\tilde{\chi}$ via $\chi = F_\chi\tilde{\chi}$
and $\kappa = (F_\pi/F_\chi)^2$.
Near chiral symmetry restoration the quarkonium component of the
dilaton field becomes a scalar mode which forms with pions an O(4)
quartet~\cite{vanKolck}. This can be formulated by making a transformation
of a non-linear chiral Lagrangian to a linear basis exploiting the dilaton limit.

The linearized Lagrangian with nucleons in the ``naive'' assignment
includes terms which generate singularities in chiral symmetric phase. 
Assuming that nature disallows any singularities, 
we require that they be absent in the Lagrangian.
This leads to $\kappa = 1$ and the Yukawa couplings (nucleon axial and vector
charges) $g_A = g_V$.
A special value, $g_V=1$, is achieved as a fixed point of the
renormalization group equations (RGEs) formulated in the chiral perturbation
theory (ChPT) with HLS when one approaches chiral restoration
from the low density or temperature side~\cite{PLRS}.
Thus, we adopt the dilaton limit as $\kappa = g_A = g_V = 1$.
In fact $g_V = g_A = 1$ recovers the large $N_c$
algebraic sum rules shown in~\cite{vanKolck}.

A noteworthy feature of the dilaton-limit Lagrangian is that the vector mesons 
decouple from the nucleons while the pion-nucleon coupling remains.
This has two striking new predictions. Taking the dilaton limit drives the Yukawa 
interaction to vanish as $g_{VN}^2=(g\,(1-g_V))^2\rightarrow 0$ for 
$V=\rho, \omega$ for any finite value of the HLS gauge coupling $g$. 
In nuclear forces, what is effective is the ratio $g_{VN}^2/m_V^2$ which goes 
as $(1-g_V)^2$. This means that (1) the two-body repulsion which holds two 
nucleons apart at short distance will be suppressed in dense medium and 
(2) the symmetry energy $S_{\rm sym}\propto g_{\rho N}^2$ will also 
get suppressed. Consequently, the EoS at some high density 
approaching the dilaton limit will become softer {\it even without any exotic 
happenings such as kaon condensation or strange quark matter}.
An interesting possibility is that our mechanism could accommodate an exotica-free 
nucleon-only EoS (e.g. AP4 in Fig. 3 of Ref.~\cite{ozel}) with a requisite 
softening at higher density that could be compatible with the 
$1.97 \pm 0.04\,M_\odot$ neutron star data~\cite{2solarmass}.

In the present scheme, the shortest-range component of the three-body forces 
also vanishes in the dilaton limit. The one-pion exchange three-body force 
involving a contact two-body force will also get suppressed as 
$\sim g_{\omega N}^2$. Thus only the longest-range two-pion exchange three-body 
forces will remain operative at large density in compact stars. How this 
intricate mechanism affects the EoS at high density is a challenging issue to 
resolve.

The dilaton limit is unchanged by the mirror baryons and therefore 
similar phenomenological consequences are expected.
Furthermore, the ChPT with either chirality assignment yields the dilaton limit
as an infrared fixed point of the coupled RGEs~\cite{PLRS}.
This could be protected at quantum level by mended symmetry, which is the 
algebraic consequence of spontaneous broken chiral symmetry~\cite{weinberg}, 
and indeed becomes manifest when the dilaton limit sets in~\cite{SLPR}.
How large is $m_0$ at the chiral symmetry restoration? 
A rough estimate can be made from thermodynamic considerations and the gluon
condensate calculated on a lattice in the presence of
dynamical quarks~\cite{miller}.
It turns out to be $m_0 = 210$ MeV and this is in agreement with the estimate 
made in vacuum phenomenology~\cite{PLRS}.

\section{Conclusions and remarks}

We have presented how an effective theory near chiral symmetry restoration
emerges from the dilaton-implemented HLS Lagrangian,
and discussed its phenomenological implications at high baryon density.
The soft dilaton is responsible for the spontaneous breaking of the scale
symmetry and its condensate vanishes when the chiral symmetry is restored.
In fact, topological stability of the half-skyrmion phase has been 
observed~\cite{half}. This is a strong indication that the configuration is
robust and it could be associated with the scale symmetry restoration at
high density in continuum theories.

Our main observation on the suppressed repulsive interaction is a common
feature in the two different assignments, ``naive'' and mirror, of chirality.
The nucleon mass near chiral symmetry restoration exhibits a striking
difference in the two scenarios. 
How the suppression of the repulsion at the dilaton limit -- which seems to be 
universal independent of the assignments but may manifest itself differently 
in the two cases -- will affect the EoS for compact stars is an interesting 
question to investigate.

In the scalar sector of low-mass hadrons, scalar quarkonium, tetra-quark 
states~\cite{jaffe} and glueballs are all mixed. In a hot/dense medium
the mixing depends on temperature and density and a multiple level-crossing
among them will be expected.
It is worth noting that
using a toy model for constituent quarks and gluons implementing chiral and
scale symmetry breaking a large sigma-meson mass $m_\sigma \sim 1$ GeV in 
matter-free space is favored along with the lattice observation of the thermal 
gluon condensate~\cite{cdm}, which is a conceivable scenario known from the 
phenomenology at zero temperature and density.

\subsection*{Acknowledgments}

I am grateful for fruitful collaboration with H.~K.~Lee, 
W.-G.~Paeng and M.~Rho.
Partial support by the Hessian
LOEWE initiative through the Helmholtz International
Center for FAIR (HIC for FAIR) is acknowledged.



\begin{thebibliography}{99}
\bibitem{walecka}
  B.~D.~Serot and J.~D.~Walecka,
  Int.\ J.\ Mod.\ Phys.\  E {\bf 6}, 515 (1997).

\bibitem{schechter}
  J.~Schechter,
  Phys.\ Rev.\  D {\bf 21}, 3393 (1980).

\bibitem{miransky}
  V.~A.~Miransky and V.~P.~Gusynin,
  Prog.\ Theor.\ Phys.\  {\bf 81}, 426 (1989).

\bibitem{LeeRho}
  H.~K.~Lee and M.~Rho,
  Nucl.\ Phys.\  A {\bf 829}, 76 (2009).

\bibitem{half}
  B.~Y.~Park, D.~P.~Min, M.~Rho and V.~Vento,
  Nucl.\ Phys.\  A {\bf 707}, 381 (2002);
  H.~J.~Lee, B.~Y.~Park, D.~P.~Min, M.~Rho and V.~Vento,
  Nucl.\ Phys.\  A {\bf 723}, 427 (2003);
  M.~Rho,  arXiv:0711.3895 [nucl-th].

\bibitem{dk}
  C.~E.~Detar and T.~Kunihiro,
  Phys.\ Rev.\  D {\bf 39}, 2805 (1989).

\bibitem{mirror}
  D.~Jido, M.~Oka and A.~Hosaka,
  Prog.\ Theor.\ Phys.\  {\bf 106}, 873 (2001).

\bibitem{SLPR}
  C.~Sasaki, H.~K.~Lee, W.~-G.~Paeng, M.~Rho,
  Phys.\ Rev.\  {\bf D84}, 034011 (2011).

\bibitem{hls}
  M.~Bando, T.~Kugo and K.~Yamawaki,
  Phys.\ Rept.\  {\bf 164}, 217 (1988);
  M.~Harada and K.~Yamawaki, Phys.\ Rept.\  {\bf 381}, 1 (2003).

\bibitem{vanKolck}
  S.~R.~Beane and U.~van Kolck,
  Phys.\ Lett.\  B {\bf 328}, 137 (1994).

\bibitem{PLRS}
  W.~-G.~Paeng, H.~K.~Lee, M.~Rho, C.~Sasaki,
  [arXiv:1109.5431 [hep-ph]].

\bibitem{ozel}
  F.~Ozel, G.~Baym and T.~Guver,
  Phys.\ Rev.\  D {\bf 82}, 101301 (2010).

\bibitem{2solarmass}
  P.~Demorest, T.~Pennucci, S.~Ransom, M.~Roberts and J.~Hessels,
  Nature {\bf 467}, 1081 (2010).

\bibitem{weinberg}
  S.~Weinberg,
  Phys.\ Rev.\ Lett.\  {\bf 65}, 1177 (1990).

\bibitem{miller} 
  D.~E.~Miller,
  Phys.\ Rept.\  {\bf 443}, 55 (2007).

\bibitem{jaffe}
  R.~L.~Jaffe,
  Phys.\ Rev.\  {\bf D15}, 267 (1977);
  Phys.\ Rev.\  {\bf D15}, 281 (1977).

\bibitem{cdm}
  C.~Sasaki, I.~Mishustin,
  [arXiv:1110.3498 [hep-ph]].

\end{thebibliography}
\end{document}